%% file: angle-monotone.tex
\documentclass{llncs}

\usepackage{makeidx}
\usepackage{hyperref}
\usepackage{graphicx}
\graphicspath{{./figs/}}
\DeclareGraphicsExtensions{.pdf,.png,.jpg}
\usepackage[linesnumbered,ruled, noline,noend]{algorithm2e}
\SetKwRepeat{Do}{do}{while}%

\usepackage{verbatim}

\usepackage{hyphenat}

\usepackage{amssymb}

\usepackage{color}

\usepackage{hyperref}
\usepackage{cite}
\usepackage{url}
\newtheorem{observation}{Observation}

\definecolor{myred}{rgb}{1.0,0.0,0.0}
\newcommand{\remove}[1]{{}}
\renewcommand{\note}[1]{{\color{myred} #1}}
\newcommand{\bluenote}[1]{{\color{blue} #1}}
\newcommand{\changed}[1]{{#1}}
\newcommand{\newstuff}[1]{{#1}}
\newcommand{\oldstuff}[1]{{}}

\newcommand{\hts}{half-\ensuremath{\theta_6}-graph\xspace}

\usepackage{tikz}
\usepackage{tkz-euclide}
\usetikzlibrary{shapes,arrows,shadows,positioning,matrix,patterns,decorations.markings,calc,intersections}
\usetkzobj{all}
\definecolor{red}{rgb}{1.0,0.0,0.0}
\definecolor{purple}{cmyk}{ .0, 0.27, 0.94, .15}

\input{figures.tex}

\title{Gabriel Triangulations and Angle-Monotone Graphs: Local Routing and Recognition}

\author{%
Nicolas Bonichon\inst{1}
\and
Prosenjit Bose\inst{2}
\and
Paz Carmi\inst{3}
\and\\
Irina Kostitsyna\inst{4}
\and 
Anna Lubiw\inst{5}
\and
Sander Verdonschot\inst{6}
}

\institute{LaBRI, Univ. Bordeaux, France.
\email{bonichon@labri.fr}
\and
 School of Computer Science, Carleton University, Ottawa, Canada. 
              \email{jit@scs.carleton.ca}
\and
Department of Computer Science, Ben-Gurion University of the Negev, Israel. 
\email{carmip@cs.bgu.ac.il}
\and
Universit\'e Libre de Bruxelles.
\email{irina.kostitsyna@ulb.ac.be}
\and
David R. Cheriton School of Computer Science, University of Waterloo, Canada.
\email{alubiw@uwaterloo.ca}
\and
School of Electrical Engineering and Computer Science, University of Ottawa, Ottawa, Canada. 
\email{sander@cg.scs.carleton.ca}
}

\tocauthor{Nicolas Bonichon, Prosenjit Bose, Paz Carmi, Irina Kostitsyna, Anna Lubiw, and Sander Verdonschot}
\date{~}

\begin{document}

\maketitle

\begin{abstract}
A geometric graph is \emph{angle-monotone} 
if every pair of vertices has a path between them that---after some rotation---is $x$- and $y$-monotone.  
Angle-monotone graphs are $\sqrt 2$-spanners and they are increasing\hyp chord graphs. 
Dehkordi, Frati, and Gudmundsson introduced angle-monotone graphs in 2014 and proved that Gabriel triangulations are angle\hyp monotone graphs.
We give a polynomial time algorithm to recognize angle-monotone geometric graphs.
We prove that every point set has a plane geometric graph that is \emph{generalized angle-monotone}---specifically, we prove that the \hts is generalized angle-monotone.
We give a local routing algorithm for Gabriel triangulations that finds a path from any vertex $s$ to any vertex $t$ whose length is within $1 + \sqrt 2$ times  the Euclidean distance from $s$ to $t$.  
Finally, we prove some lower bounds and limits on local routing algorithms on Gabriel triangulations.
\end{abstract}

\section{Introduction}
\label{sec:introduction}
\input introduction.tex

\section{Recognizing Angle-monotone Graphs}
\label{sec:recognition}
\input recognition.tex

\section{A Class of Generalized Angle-Monotone Graphs}
\label{sec:gen}
\input half-theta-6.tex

\section{Local Routing in Gabriel Triangulations}
\label{sec:routing}
\input routing.tex


\section{Conclusions}
\label{sec:conclusions}
\input{conclusions.tex}

\section*{Acknowledgements}

This work was begun at the CMO-BIRS Workshop on Searching and Routing in Discrete and Continuous Domains,
October 11--16, 2015.  We thank the other participants of the workshop for many good ideas and stimulating discussions.  \changed{We thank an anonymous referee for helpful comments.}

Funding acknowledgements: 
A.L.~thanks NSERC (Natural Sciences and Engineering Council of Canada).
S.V.~thanks NSERC and the Ontario Ministry of Research and Innovation.
N.B.~thanks French National Research Agency (ANR) in the frame of the ``Investments for the future'' Programme IdEx Bordeaux - CPU (ANR-10-IDEX-03-02).
I.K.~was supported in part by the NWO under project no. 612.001.106, and by F.R.S.-FNRS.

\bibliographystyle{splncs03}
\bibliography{angle-monotone}

\appendix
\input{appendix.tex}

\end{document}

%% file: figures.tex
\newcommand{\nodeList}{{0/250/p0/$s$/below left},{338/250/t/$t$/},{60/210/p1/$p_1$/below left},{108/277/q2/$q_2$/above},{144/187/p2/$p_2$/},{178/235/p3/$p_3$/below},{243/279/p4/$p_4$/above},{237/184/q4/$q_4$/},{295/239/q5/$q_5$/},{305/307/p5/$p_5$/above},{189/312/q1p/$q_3$/above},{54/309/q1/$q_1$/above}}
\newcommand{\routingPath}{{p0/p1/q1/1/below/90/0/1},{p1/p2/q2/0/above/-140/1/1},{p2/p3/q2/0/above/-140/1/1},{p3/p4/q4/0/above/-140/1/1},{p4/p5/q5/0/above/-140/1/1},{p5/t/q5/0/above/-140/1/1}}
\newcommand{\otherTriangles}{{q1/p1/q2},{q2/p3/q1p},{p3/p4/q1p},{p4/q4/q5}}
\newcommand{\myAngle}{90}

\newcommand{\clipExample}{\clip (-0.3,1.3) rectangle (3.8,3.75);}
\newcommand{\clipExampleLB}{\clip (-0.3,1.8) rectangle (3.8,5.5);}
\newcommand{\blue}{black}
\tikzstyle{path}=[color=blue,line width=2.5pt]
\tikzstyle{pathBound}=[color=red,line width=2.pt,style=dotted]
\tikzstyle{emptyCircle}=[color=gray]
\tikzstyle{triangles}=[color=\blue]

\newcommand{\tkzComputeAllPoints}{
\tkzDefMidPoint(p0,p0)\tkzGetPoint{s}

\foreach \A/\B/\C  in \otherTriangles
{\tkzCircumCenter(\A,\B,\C)\tkzGetPoint{G\A\B\C} }

\foreach \A/\B/\C/\cw/\cpos/\cposangle/\cnum/\ctype  in \routingPath
  {    \tkzDefMidPoint(\A,\B)\tkzGetPoint{mid\A\B}
       \tkzDefLine[parallel=through \A](s,t)\tkzGetPoint{para\A}
       \tkzDefLine[orthogonal=through mid\A\B](\A,\B) \tkzGetPoint{ortho\A\B}
       \tkzInterLL(mid\A\B,ortho\A\B)(\A,para\A) \tkzGetPoint{Gp\A\B\C}
  \tkzCircumCenter(\A,\B,\C)\tkzGetPoint{G\A\B\C}
  \tkzDefMidPoint(\A,\A)\tkzGetPoint{primep\cnum}
}

\foreach \A/\B/\C/\cw/\cpos/\cposangle/\cnum/\ctype  in \routingPath
{ \tkzDefPointBy[projection=onto s--t](primep\cnum) \tkzGetPoint{sp\cnum}
\tkzDefPointBy[projection=onto s--t](primep\cnum) \tkzGetPoint{s\cnum}}

\tkzDefPointBy[translation= from s1 to p1](s) \tkzGetPoint{s0}
\tkzDefPointBy[projection=onto p1--s1](p2) \tkzGetPoint{s1}
}

\newcommand{\figexampleRouting}{
\begin{tikzpicture}[scale=2.0]
\clipExample

\foreach \x\y/\name/\etiq/\pos  in \nodeList
  {\tkzDefPoint(\x/100,\y/100){\name}}

\tkzComputeAllPoints

\tkzDefLine[parallel=through p1](s,t) \tkzGetPoint{pp1}
\tkzDefLine[parallel=through p2](s,t) \tkzGetPoint{pp2}
\tkzDefLine[parallel=through p3](s,t) \tkzGetPoint{pp3}
\tkzDefLine[parallel=through p4](s,t) \tkzGetPoint{pp4}
\tkzDefLine[parallel=through p5](s,t) \tkzGetPoint{pp5}
\tkzDefLine[parallel=through t](s,t) \tkzGetPoint{pp6}

\tkzDefPointBy[rotation=center s angle 180+\myAngle](t) \tkzGetPoint{tm90}
\tkzDefPointBy[rotation=center s angle 180-\myAngle](t) \tkzGetPoint{tp90}
\tkzDefPointBy[rotation=center s angle -\myAngle/2](t) \tkzGetPoint{tm45}
\tkzDefPointBy[rotation=center s angle \myAngle/2](t) \tkzGetPoint{tp45}

\tkzDefLine[parallel=through p0](s,tm45) \tkzGetPoint{pq0}
\tkzDefLine[parallel=through p1](s,tm45) \tkzGetPoint{pq1}
\tkzDefLine[parallel=through p2](s,tm90) \tkzGetPoint{pq2}
\tkzDefLine[parallel=through p3](s,tp45) \tkzGetPoint{pq3}
\tkzDefLine[parallel=through p4](s,tp45) \tkzGetPoint{pq4}
\tkzDefLine[parallel=through p5](s,tp90) \tkzGetPoint{pq5}

\tkzInterLL(p0,pq0)(pp1,p1) \tkzGetPoint{p01}
\tkzInterLL(p1,pq1)(pp2,p2) \tkzGetPoint{p12}
\tkzInterLL(p2,pq2)(pp3,p3) \tkzGetPoint{p23}
\tkzInterLL(p3,pq3)(pp4,p4) \tkzGetPoint{p34}
\tkzInterLL(p4,pq4)(pp5,p5) \tkzGetPoint{p45}
\tkzInterLL(p5,pq5)(pp6,t) \tkzGetPoint{p56}

\tkzDrawSegment[dashed](s,t)

\foreach \A/\B/\C  in \otherTriangles
{\tkzCircumCenter(\A,\B,\C)\tkzGetPoint{G\A\B\C}
   \tkzDrawPolygon[triangles, color=gray](\A,\B,\C)
   }

\foreach \A/\B/\C/\cw/\cpos/\cposangle/\cnum/\ctype  in \routingPath
  {\tkzCircumCenter(\A,\B,\C)\tkzGetPoint{G\A\B\C}
   \tkzDrawPolygon[triangles](\A,\B,\C)
   \tkzDrawSegment[path, line width=2pt](\B,\A)}

\tkzDrawSegments[pathBound](p0,p01  p01,p1)
\tkzDrawSegments[pathBound](p1,p12  p12,p2)
\tkzDrawSegments[pathBound](p2,p23  p23,p3)
\tkzDrawSegments[pathBound](p3,p34  p34,p4)
\tkzDrawSegments[pathBound](p4,p45  p45,p5)
\tkzDrawSegments[pathBound](p5,p56  p56,t)

\foreach \x\y/\name/\etiq/\pos  in \nodeList
  { \tkzDrawPoint[fill=white,size=10](\name)
    \tkzLabelPoint[\pos](\name){\etiq}
  }
\tkzDrawPoint[size=10,fill=blue](s)
\tkzDrawPoint[size=10,fill=red](t)
\end{tikzpicture}
}

\newcommand{\figexampleRoutingGeneralized}{
\renewcommand{\nodeList}{{0/250/p0/$s$/below left},{348/250/t/$t$/},{60/210/p1/$p_1$/below left},{108/257/q2/$q_2$/above},{154/187/p2/$p_2$/},{148/235/p3/$p_3$/below},{243/279/p4/$p_4$/above},{237/184/q4/$q_4$/},{295/239/q5/$q_5$/},{365/327/p5/$p_5$/above},{189/312/q1p/$q_3$/above},{54/309/q1/$q_1$/above}}
\renewcommand{\routingPath}{{p0/p1/q1/1/below/90/0/1},{p1/p2/q2/0/above/-140/1/1},{p2/p3/q2/0/above/-140/1/1},{p3/p4/q4/0/above/-140/1/1},{p4/p5/q5/0/above/-140/1/1},{p5/t/q5/0/above/-140/1/1}}
\renewcommand{\otherTriangles}{{q1/p1/q2},{q2/p3/q1p},{p3/p4/q1p},{p4/q4/q5}}
\renewcommand{\myAngle}{110}
\figexampleRouting
}

\newcommand{\figLBDel}{
\renewcommand{\nodeList}{{0.7/250/p0/$s$/left},{100/250/t/$t$/below},{4/233/q1/$q_1$/below left},{4/268/p1/$p_1$/above left},{96/250/q2/$q_2$/below left},{251/310/p2/$p_2$/},{221/286/p3//},{168/261/p5//},{192/271/p4//}}

\renewcommand{\routingPath}{{p0/p1/q1/1/below/90/0/1},{p1/p2/q2/0/above/-140/1/1},{p2/p3/q2/0/above/-140/1/1},{p3/p4/q2/0/above/-140/1/1},{p4/p5/q2/0/above/-140/1/1},{p5/t/q2/0/above/-140/1/1}}
\renewcommand{\otherTriangles}{{q1/p1/q2}}
\begin{tikzpicture}[scale=2.6]
\clip (-0.8,1.3) rectangle (3.5,3.5);

\foreach \x\y/\name/\etiq/\pos  in \nodeList
  {\tkzDefPoint(\x/100,\y/100){\name}}

\tkzComputeAllPoints

   \tkzDrawCircle[emptyCircle](Gp1p2q2,p1)
   \tkzDrawCircle[emptyCircle](Gp0p1q1,p1)

\tkzDrawSegment[dashed](s,t)

\foreach \A/\B/\C  in \otherTriangles
{\tkzCircumCenter(\A,\B,\C)\tkzGetPoint{G\A\B\C}
   \tkzDrawPolygon[triangles, color=gray](\A,\B,\C)
   }

\foreach \A/\B/\C/\cw/\cpos/\cposangle/\cnum/\ctype  in \routingPath
  {\tkzCircumCenter(\A,\B,\C)\tkzGetPoint{G\A\B\C}
   \tkzDrawPolygon[triangles](\A,\B,\C)

   \tkzDrawSegment[path, line width=2pt](\B,\A)}

\foreach \x\y/\name/\etiq/\pos  in \nodeList
  { \tkzDrawPoint[fill=white,size=10](\name)
    \tkzLabelPoint[\pos](\name){\etiq}
  }
\tkzDrawPoint[size=10,fill=blue](s)
\tkzDrawPoint[size=10,fill=red](t)
\end{tikzpicture}
}

\newcommand{\figLBAlgo}{
\begin{tikzpicture}[scale=1.5]
\pgfmathsetmacro{\xAxis}{250}
\renewcommand{\nodeList}{{0/250/p0/$s$/left},{240/250/t/$t$/},{30/220/q1/$q_1$/},{30/280/p1/$p_1$/above left},{60/246/q2/$q_2$/},{236/486/p2/$p_2$/}}

\renewcommand{\routingPath}{{p0/p1/q1/1/below/90/0/1},{p1/p2/q2/0/above/-140/1/1},{p2/t/q2/0/above/-140/1/1}}
\renewcommand{\otherTriangles}{{q1/p1/q2}}

\clipExampleLB
\foreach \x\y/\name/\etiq/\pos  in \nodeList
  {\tkzDefPoint(\x/100,\y/100){\name}}

\tkzComputeAllPoints

\tkzDrawSegment[dashed](s,t)

\foreach \A/\B/\C  in \otherTriangles
  {\tkzDrawPolygon[triangles, color=gray](\A,\B,\C)}

\foreach \A/\B/\C/\cw/\cpos/\cposangle/\cnum/\ctype  in \routingPath
  {\tkzDrawPolygon[triangles](\A,\B,\C)
   \tkzDrawSegment[path, line width=2pt](\B,\A)}

\foreach \x\y/\name/\etiq/\pos  in \nodeList
  { \tkzDrawPoint[fill=white,size=10](\name)
    \tkzLabelPoint[\pos](\name){\etiq}}

\tkzDrawPoint[size=10,fill=blue](s)
\tkzDrawPoint[size=10,fill=red](t)
\end{tikzpicture}
}

\newcommand{\figLBAlgoGeneralized}{
\begin{tikzpicture}[scale=1.5]
\pgfmathsetmacro{\xAxis}{250}
\renewcommand{\nodeList}{{0/250/p0/$s$/left},{96/250/t/$t$/},{12/233/q1/$q_1$/},{12/267/p1/$p_1$/above left},{24/249/q2/$q_2$/},{191/523/p2/$p_2$/}}

\renewcommand{\routingPath}{{p0/p1/q1/1/below/90/0/1},{p1/p2/q2/0/above/-140/1/1},{p2/t/q2/0/above/-140/1/1}}
\renewcommand{\otherTriangles}{{q1/p1/q2}}

\clipExampleLB
\foreach \x\y/\name/\etiq/\pos  in \nodeList
  {\tkzDefPoint(\x/100,\y/100){\name}}

\tkzComputeAllPoints

\tkzDrawSegment[dashed](s,t)
\tkzMarkAngles[size=0.2, fill=green, opacity=0.5](p2,t,q2)
\tkzLabelAngle[pos=-0.2, above left](p2,t,q2){$\alpha$}
\foreach \A/\B/\C  in \otherTriangles
  {\tkzDrawPolygon[triangles, color=gray](\A,\B,\C)}

\foreach \A/\B/\C/\cw/\cpos/\cposangle/\cnum/\ctype  in \routingPath
  {\tkzDrawPolygon[triangles](\A,\B,\C)
   \tkzDrawSegment[path, line width=2pt](\B,\A)}

\foreach \x\y/\name/\etiq/\pos  in \nodeList
  { \tkzDrawPoint[fill=white,size=10](\name)
    \tkzLabelPoint[\pos](\name){\etiq}}

\tkzDrawPoint[size=10,fill=blue](s)
\tkzDrawPoint[size=10,fill=red](t)
\end{tikzpicture}
}

\newcommand{\figLB}{
\begin{tikzpicture}[x=2.0cm,y=2.0cm,scale=1.9]
\clip(-0.2,-1.0) rectangle (2, 1.1);
\pgfmathsetmacro{\myLen}{0.5}
\pgfmathsetmacro{\myEps}{0.1}

\tkzDefPoint(\myLen, 0){O}
\tkzDefPoint(-\myLen, 0){sp}
\tkzDefPoint(\myLen, \myLen){q}
\tkzDefPoint(\myLen, -\myLen){qp}

\tkzDefPoint(\myLen-\myEps, 0){Obis}
\tkzDefPoint(\myLen-\myEps, \myLen){qbis}
\tkzDefPoint(\myLen-\myEps, -\myLen){qpbis}
\tkzDefMidPoint(q,qbis)\tkzGetPoint{qter}
\tkzDefMidPoint(qp,qpbis)\tkzGetPoint{qpter}
\tkzDefPoint(-\myEps, 0){s}
\tkzDefPointBy[rotation= center O angle 45](qp)\tkzGetPoint{B}
\tkzDefPointBy[rotation= center B angle -90](q)\tkzGetPoint{t}
\tkzDefMidPoint(q,t)\tkzGetPoint{O2}
\tkzDefPointBy[translation= from B to O2](O2) \tkzGetPoint{A}

\tkzDrawSegment[dashed](s,t)

\tkzDrawCircle[emptyCircle](O,q)
\tkzDrawCircle[emptyCircle](O2,A)
\tkzDrawCircle[emptyCircle](Obis,qbis)

\tkzMarkAngles[size=0.2, fill=green, opacity=0.5](qp,q,B)
\tkzLabelAngle[right,pos=0.25](qp,q,B){$22.5^\circ$}
 \tkzDrawPolygon[triangles](s, qbis, qpbis)
 \tkzDrawPolygon[triangles](B, q, qp)
 \tkzDrawPolygon[triangles](B, A, t)
 \tkzDrawPolygon[triangles](B, A, q)
 \tkzDrawSegment[triangles](qbis,q)
 \tkzDrawSegment[triangles](qpbis,qp)
 \tkzDrawSegment[triangles](qbis,qpter)
 \tkzDrawSegment[triangles](qter,qp)

\tkzDrawPoint[size=10,fill=white](q)\tkzLabelPoint[above](q){$q$}
\tkzDrawPoint[size=10,fill=white](qp)\tkzLabelPoint(qp){$q'$}
\tkzDrawPoint[size=10,fill=white](qbis)
\tkzDrawPoint[size=10,fill=white](qpbis)
\tkzDrawPoint[size=10,fill=white](qpter)
\tkzDrawPoint[size=10,fill=white](qter)
\tkzDrawPoint[size=10,fill=white](B)\tkzLabelPoint[below](B){$B$}
\tkzDrawPoint[size=10,fill=white](A)\tkzLabelPoint[above right](A){$A$}
\tkzDrawPoint[size=10, fill=red](t)\tkzLabelPoint(t){$t$}
\tkzDrawPoint[size=10, fill=blue](s)\tkzLabelPoint[left](s){$s$}

\end{tikzpicture}
}

\newcommand{\figNoSA}{
\begin{tikzpicture}[scale=3]
\clip(-0.2,-1.0) rectangle (2, 1.1);

\tkzDefPoint(0.5, 0){O}
\tkzDefPoint(0.5, 0.5){q}
\tkzDefPoint(0.5, -0.5){qp}
\tkzDefPoint(1.05, -0.15){B}
\tkzDefPoint(0, 0){s}
\tkzDefPoint(1.6, 0){t}
\tkzDefPoint(1.3, 0.9){A}
\tkzDefMidPoint(A,B) \tkzGetPoint{O2}
\tkzDefPointBy[projection=onto A--q](B) 

\tkzDefLine[perpendicular=through B](s,q)\tkzGetPoint{Bpa}
\tkzDefLine[perpendicular=through t](q,A)\tkzGetPoint{tpa}
\tkzInterLL(q,A)(t,tpa) \tkzGetPoint{tp}
\tkzInterLL(s,q)(B,Bpa) \tkzGetPoint{Bp}
\tkzDrawSegment[color=gray](B,Bp)
\tkzDrawSegment[color=gray](t,tp)
\tkzMarkRightAngle[color=gray, size=0.07](s,Bp,B)
\tkzMarkRightAngle[color=gray, size=0.07](q,tp,t)

\tkzDrawSegment[dashed](s,t)

\tkzDrawCircle[emptyCircle](O,q)
\tkzDrawCircle[emptyCircle](O2,A)

\tkzDrawPolygon[triangles](s, q, qp)
\tkzDrawSegment[line width=2pt](q, qp)
 \tkzDrawPolygon[triangles](B, q, qp)
 \tkzDrawPolygon[triangles](B, A, t)
 \tkzDrawPolygon[triangles](B, A, q)

\tkzDrawPoint[size=10,fill=white](q)\tkzLabelPoint[above](q){$q$}
\tkzDrawPoint[size=10,fill=white](qp)\tkzLabelPoint(qp){$q'$}

\tkzDrawPoint[size=10,fill=white](B)\tkzLabelPoint[below](B){$B$}
\tkzDrawPoint[size=10,fill=white](A)\tkzLabelPoint[above right](A){$A$}
\tkzDrawPoint[size=10, fill=red](t)\tkzLabelPoint(t){$t$}
\tkzDrawPoint[size=10, fill=blue](s)\tkzLabelPoint[left](s){$s$}

\end{tikzpicture}
}

%% file: introduction.tex


A geometric graph has vertices that are points in the plane, and edges that are drawn as straight-line segments, with the weight of an edge being its Euclidean length.  A geometric graph need not be planar.
Geometric graphs that have relatively short paths are relevant in many applications for routing and network design, 
and have been a subject of intense research.
A main scenario is that we are given a point set and must construct a sparse geometric graph on that point set with good shortest path properties.  

If the shortest path between every pair of points has length at most $t$ times the Euclidean distance between the points, then 
the geometric graph is called a $t$-\emph{spanner}, and 
\changed{the minimum such} $t$ is called the \emph{spanning ratio}.  
Since their introduction by Paul Chew in 1986~\cite{Chew86}, spanners have been heavily studied~\cite{Spanners}. 

Besides the existence of short paths, another issue is
\emph{routing}---how to find short paths in a geometric graph.  
One goal is to find paths using \emph{local routing} where the path is found one vertex at a time using only local information about the neighbours of the current vertex plus the coordinates of the destination.
A main example of such a method is \emph{greedy routing}: from the current vertex $u$ take any edge to a vertex $v$ that is closer (in Euclidean distance) to the destination than $u$ is.
The geometric graphs for which greedy routing succeeds in finding a path are called \emph{greedy drawings}.
These have received considerable attention because of their potential ability to replace routing tables for network routing, and because of the noted conjecture of Papadimitriou and Ratajczak~\cite{Papadimitriou:2005} 
(proved in~\cite{Leighton:2010,Angelini:2009}) that every 3-connected planar graph has a greedy drawing.
One drawback is that a path found by greedy routing may be very long compared to the Euclidean distance between the endpoints.  Of course this is inevitable if the \changed{geometric} graph has large spanning ratio.

When a \changed {geometric} graph is a $t$-spanner, we can ideally hope for a local routing algorithm
that finds a path whose length is at most $k$ times the Euclidean distance between the endpoints, for some $k$,  where, of necessity, $k \ge t$.   The \changed{maximum} ratio, $k$, of path length to Euclidean distance is called the \emph{routing ratio}.
For example, the Delaunay triangulation, which is a $t$-spanner for $t \le 1.998$~\cite{xia2013stretch}, permits local routing with routing ratio $k \le 5.90$~\cite{BBCPR15}.  It is an open question whether the spanning ratio and routing ratio are equal, though there is a provable gap for $L_1$-Delaunay triangulations~\cite{BBCPR15} and TD-Delaunay triangulations~\cite{BoseTheta6journal}.
  
\smallskip\noindent{\bf Other ``good'' paths.}
Recently, a number of other notions of ``good'' paths in geometric graphs have been investigated.  
Alamdari et al.~\cite{Alamdari2013} introduced \emph{self-approaching} graphs, where any two vertices $s$ and $t$ are joined by a \emph{self-approaching path}---a path such that a point moving continuously along the path from $s$ to any intermediate destination $r$ on the path always gets closer to $r$ in Euclidean distance.  
In an \emph{increasing-chord} graph, this property also holds for the reverse path from $t$ to $s$. 
The self-approaching path property is stronger than the greedy path property in two ways:
\changed{it applies to every intermediate destination $r$, and it requires that continuous motion (not just the vertices) along the path to $r$ always gets closer to $r$.}
The significance of the stronger property is that 
self-approaching and increasing-chord graphs have bounded spanning ratios of 5.333~\cite{Icking:self-approachingcurves:1995} and 2.094~\cite{Rote:ICcurves:1994}, respectively. 
An important characterization is that a path is self-approaching if and only if at each point on the path, there is a $90^\circ$ wedge that contains the rest of the path~\cite{Icking:self-approachingcurves:1995}.  

Angelini et al.~\cite{Angelini:MonoDraw:2012} introduced  \emph{monotone drawings}, where any two vertices $s$ and $t$ are joined by a 
path that is monotone in some direction.  This is a natural desirable property, but not enough to guarantee a bounded spanning ratio. 

\smallskip\noindent{\bf Angle-monotone paths.}
In this paper we explore properties of another class of geometric graphs with good path properties.  These are the \emph{angle-monotone graphs} which were first introduced by Dehkordi, Frati, and Gudmundsson~\cite{D-Frati-G} as a tool to investigate increasing-chord graphs.  (We note that Dehkordi et al.~\cite{D-Frati-G} did not give a name to their graph class.)

\begin{figure}[htb]
\centering
\includegraphics[width=0.6\textwidth]{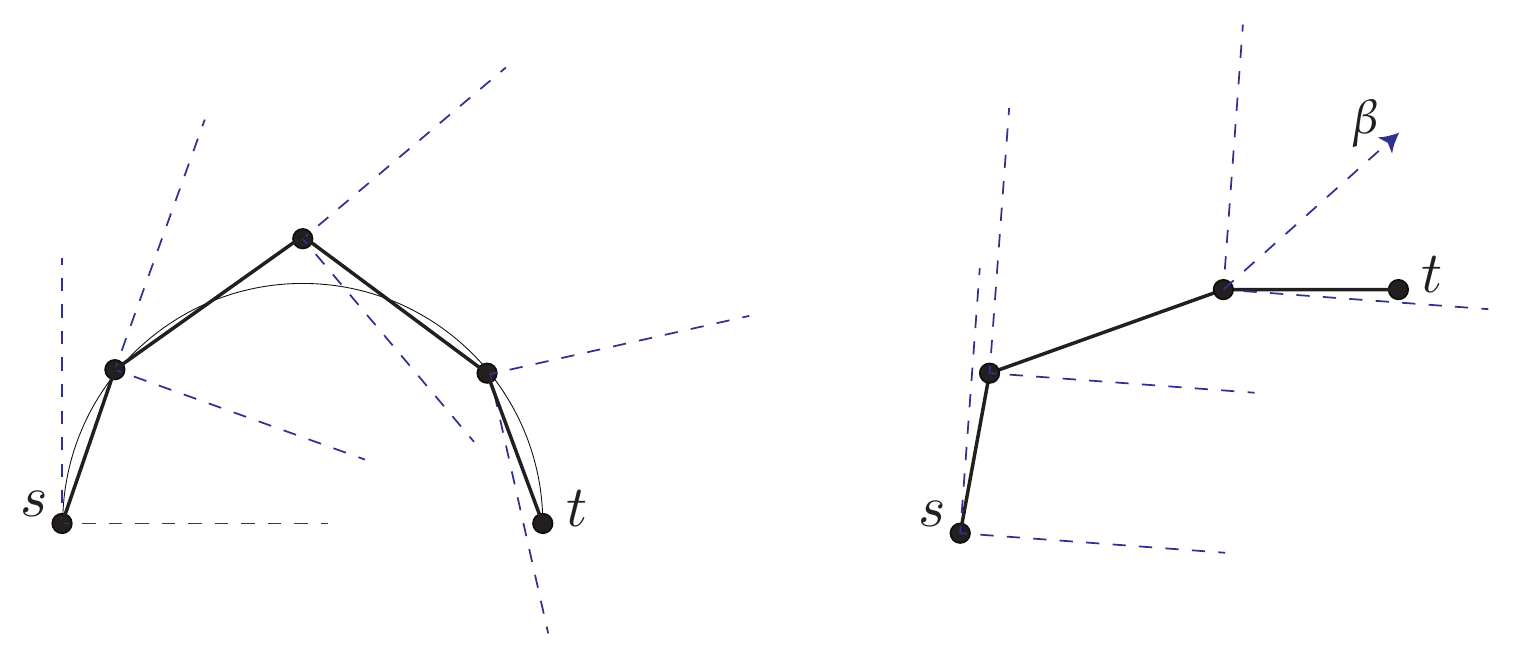}
\caption{The difference between a self-approaching $st$ path (left) with $90^\circ$ wedges each containing the rest of the path, and an angle-monotone path (right) where the $90^\circ$ wedges all have the same orientation $\beta$.}
\label{fig:SA-vs-AM}
\end{figure}

A polygonal path with vertices $v_0, v_1, \ldots, v_n$ is \emph{$\beta$-monotone} for some angle $\beta$ if the vector of every edge $(v_i, v_{i+1})$ lies in the closed $90^\circ$ wedge between $\beta - 45^\circ$ and $\beta + 45^\circ$.
(In the terminology of  Dehkordi et al.~\cite{D-Frati-G} this is a \emph{$\theta$-path}.)
In particular, an $x$-$y$-monotone path (where $x$ and $y$ coordinates are both non-decreasing along the path) is a $\beta$-monotone path for $\beta = 45^\circ$ (measured from the positive $x$-axis).
A path is \emph{angle-monotone} if there is some angle $\beta$ for which it is $\beta$-monotone.
To visualize this, note that  a path is 
angle-monotone if and only if it can be rotated to be $x$-$y$-monotone. 
An angle-monotone path is a special case of a self-approaching path where the wedges containing the rest of the path all have the same orientation.  See Figure~\ref{fig:SA-vs-AM}.  This implies that an angle-monotone path is also angle-monotone when traversed in the other direction, and thus, has the increasing-chord property.  
Observe that angle-monotone paths have spanning ratio $\sqrt 2$---this is because $x$-$y$-monotone paths do.

A geometric graph is \emph{angle-monotone} if for every pair of vertices $u$, $v$, there is an angle-monotone path from $u$ to $v$.  
Note that the angle $\beta$ may be different for different pairs $u,v$.
Dehkhori et al.~\cite{D-Frati-G} introduced angle-monotone graphs, and proved that they include the class of Gabriel triangulations (triangulations with no obtuse angle).
Their main goal was to prove 
that any set of $n$ points in the plane has a planar increasing-chord graph with $O(n)$ Steiner points and $O(n)$ edges.  Given their result that Gabriel graphs are increasing chord, this follows from a result of Bern et al.~\cite{bern1990provably} that any point set can be augmented with $O(n)$ points to a point set whose Delaunay triangulation is Gabriel.

The notion of angle-monotone graphs can be generalized to wedges of angle $\gamma$ different from $90^\circ$.
(A precise definition is given below.)
We call these \emph{angle-monotone graphs with width $\gamma$}, or \emph{generalized angle-monotone graphs.}  For  $\gamma < 180^\circ$, they still have bounded spanning ratios.

\medskip\noindent{\bf Results.}  The main themes we explore are:  Which geometric graphs are angle-monotone?  Can we create a sparse (generalized) angle-monotone graph on any given point set?  Do angle-monotone graphs permit local routing?

Our first main result is a polynomial time algorithm to test if a geometric graph is angle-monotone. 
This is significant because it is not known whether increasing chord graphs can be recognized in polynomial time (or whether the problem is NP-hard).  
Our algorithm extends to generalized angle-monotone graphs for any width $\gamma < 180^\circ$.

Our next result is that for any point set in the plane, there is a plane geometric graph on that point set that is angle-monotone with width $120^\circ$.  In particular, we prove that the \emph{\hts} has this property.
Width $90^\circ$ cannot always be achieved because it would imply spanning ratio $\sqrt 2$ which is known to be impossible for some point sets\changed{, as discussed below under Further Background.}

The rest of the paper is about local routing algorithms, where we concentrate on a subclass of angle-monotone graphs, namely the Gabriel triangulations.
We give a local routing algorithm for Gabriel triangulations that achieves routing ratio $1 + \sqrt 2 \thickapprox 2.41$.  
This is better than the best known routing ratio for Delaunay triangulations of 5.90~\cite{BBCPR15}.
Also, our algorithm is simpler.
The algorithm \changed{succeeds, i.e.~finds a path to the destination, for} 
any triangulation, and we prove that the algorithm has a bounded routing ratio for triangulations with maximum angle less than $120^\circ$.
\changed{For Delaunay triangulations, we prove a lower bound on the routing ratio of 5.07, but leave as an open question whether the algorithm ever does worse.}
Finally, we give some lower bounds on the routing ratio of local routing algorithms on Gabriel triangulations, and we prove that no local routing algorithm on Gabriel triangulations can find self-approaching paths.


As is clear from this outline, we leave many interesting open questions, some of which are listed in the Conclusions section.


\medskip\noindent{\bf Further Background.}
\remove{
The Delaunay triangulation is the ``best'' triangulation of a set of points for many purposes.  The standard \emph{Delaunay triangulation} is defined to have an edge  $uv$ if there is a disc with $u$ and $v$ on its boundary that contains no other point.
Replacing the disc in this definition by other convex shapes leads to ``generalized Delaunay triangulations'', for example the $L_1$ Delaunay triangulation uses a square with sides of slope $1$ and $-1$.
The standard Delaunay triangulation is a $1.998$-spanner~\cite{xia2013stretch} and permits local routing with routing ratio $k \le 5.90$~\cite{BBCPR15}  using a generalization of Chew's routing algorithm for $L_1$ Delaunay triangulations~\cite{Chew86}.  For more information on local routing in (generalized) Delaunay triangulations and for results on the gaps between the spanning ratio and the routing ratio, see~\cite{BBCPR15}.
}
The standard Delaunay triangulation is not self\hyp approaching in general~\cite{Alamdari2013}, and therefore not angle-monotone.  

The \emph{Gabriel graph} of point set $P$ is a graph in which for every edge $(u,v)$ the circle with diameter $uv$ contains no points of $P$. A Gabriel graph that is a triangulation is called a \emph{Gabriel triangulation}.  Any Gabriel triangulation is a Delaunay triangulation.
Observe that a triangulation is Gabriel if and only if it has no obtuse angles.  \changed{Not every point set has a Gabriel triangulation, e.g.~three points forming an obtuse triangle.}

There are several results on constructing self-approaching/increasing-chord graphs on a given set of points. 
Alamdari et al.~\cite{Alamdari2013} constructed an increasing chord network of linear size using Steiner points, and Dehkordi et al.~\cite{D-Frati-G} improved this to a plane network.  
It is an open question whether every point set admits a plane increasing-chord graph without adding Steiner points. 
However, for the more restrictive case of angle-monotone graphs, the answer is no: 
any angle-monotone graph has spanning ratio $\sqrt 2$ but there is a point set (specifically, the vertices of a regular 23-gon) for which any planar geometric graph has spanning ratio at least 1.4308~\cite{dumitrescu2015lower}.   An earlier example was given by Mulzer~\cite{mulzer}.


\medskip\noindent{\bf Preliminaries and Definitions.}
A polygonal path with vertices $v_0, v_1, \ldots, v_n$ is \emph{$\beta$-monotone with width $\gamma$} for some angles $\beta$ and $\gamma$ with $\gamma < 180^\circ$ if the vector of every edge $(v_i, v_{i+1})$ lies in the closed wedge of angle $\gamma$ between $\beta - {\gamma \over 2}$ and $\beta + {\gamma \over 2}$. 
When we have no need to specify $\beta$, we say that the path is  
\emph{angle-monotone with width $\gamma$}, or ``generalized angle-monotone''.  
A path that is generalized angle-monotone is a \emph{generalized self-approaching path}~\cite{Aichholzer:genSAcurves:2001} and thus has bounded spanning ratio depending on $\gamma$~\cite{Aichholzer:genSAcurves:2001}.
But in fact, we can do better: 

\begin{observation} \label{obs:ratio} [proof in Appendix~\ref{sec:appendix-intro}]
The spanning ratio of an angle-monotone path with width $\gamma < 180^\circ$ is at most $1/\cos{\gamma \over 2}$. 
\end{observation}

A geometric graph is \emph{angle-monotone with width $\gamma$} if for every pair of vertices $u$, $v$, there is an angle-monotone path with width $\gamma$ from $u$ to $v$.  When we have no need to specify $\gamma$, we say that the graph is ``generalized angle-monotone''. 

Note that in an angle-monotone path (with width $90^\circ$) the distances from $v_0$ to later vertices form an increasing sequence.   Furthermore,  any $\beta$-monotone path from $u$ to $v$ lies in a rectangle with $u$ and $v$ at opposite corners and with two sides at angles $\beta \pm 45^\circ$, and the union of such rectangles over all \changed{$\beta \in [0,360^\circ)$} forms the disc with diameter $uv$.  \changed{(See Figure~\ref{fig:lemma1} in Appendix.)}
This implies:

\begin{lemma}
Any angle-monotone path from $u$ to $v$ lies inside the disc with diameter $uv$.
\label{lemma:disc}
\end{lemma}

%

%% file: recognition.tex


In this section we give an $O(n m^2)$ time algorithm to test if a geometric graph with $n$ vertices and $m$ edges is angle-monotone.
The idea is to look for angle-monotone paths from a node $s$ to all other nodes, and then repeat over all choices of $s$.  
For a given source vertex $s$, the algorithm explores nodes $u$ in non-decreasing order of their distance from $s$.
At each vertex $u$ we store information to capture all the possible angles $\beta$ for which there is a $\beta$-monotone path from $s$ to $u$.  We show how to propagate this information along an edge from $u$ to $v$.

We begin with some notation.
We will measure angles counterclockwise from the positive $x$-axis, modulo $360^\circ$.
To any ordered pair $u,v$ of vertices (points) of our geometric graph  
we associate the vector $v-u$ and we denote its angle by $\alpha(u,v)$. 
If $S$ is a set of angles that lie within a wedge of angle less than $180^\circ$, then we define the \emph{minimum} of $S$ to be the most clockwise angle, and the \emph{maximum} of $S$ to be the most counter-clockwise angle.
More formally, $\alpha$ is the \emph{minimum} of $S$ if for any other $\beta \in S$, $\beta - \alpha \in [0,180^\circ)$, and similarly for \emph{maximum}.

 
Although there may be exponentially many angle-monotone paths from $s$ to $u$, each such path has two extreme edges.  More precisely, if $P$ is an angle-monotone path from $s$ to $u$, then the angles, $\alpha(e), e \in P$, lie in a $90^\circ$ wedge, and so this set has a minimum and maximum that differ by at most $90^\circ$.  We will store a list of all such min-max pairs with vertex $u$.  Each pair defines a wedge of at most $90^\circ$.
Since each pair is defined by two edges, there are at most $O(m^2)$ such pairs (though we will show below that we only need to store $O(m)$ of them). 

The algorithm starts off by looking at every edge $(s,u)$ and adding the pair $(\alpha(s,u), \alpha(s,u))$ to $u$'s list. 
Then the algorithm explores vertices $u \ne s$ in non-decreasing order of their distance from $s$.
To explore vertex $u$, consider each edge $(u,v)$ and each pair $(\alpha(e), \alpha(f))$ stored with $u$, and update the list of pairs for vertex $v$ as follows.   If $\alpha(u,v)$ is within $90^\circ$ of $\alpha(e)$ and within $90^\circ$ of $\alpha(f)$ then add to $v$'s list the pair $(\min\{ \alpha(u,v), \alpha(e)\}, \max\{\alpha(u,v), \alpha(f)\})$. 


If ever the algorithm tries to explore a vertex that has no pairs stored with it, then 
halt---the graph is not angle-monotone.
To justify correctness we prove:

\begin{lemma}
When the algorithm has explored all the vertices closer to $s$ than $v$, then there exists an angle-monotone path from $s$ to $v$ with extreme edges $e$ and $f$ if and only if the pair $(\alpha(e), \alpha(f))$ is in $v$'s list.
\end{lemma}
\begin{proof}
The proof is by induction on the distance from $s$ to $v$.

For the ``only if'' direction, let $P$ be an angle-monotone path from $s$ to $v$ with extreme edges $e$ and $f$, and let $u$ be the penultimate vertex of $P$.
The subpath of $P$ from $s$ to $u$ is an angle-monotone path.  Suppose its extreme edges are $e'$ and $f'$ where $e=e'$ or $f=f'$ or both.  
Now, $u$ is closer to $s$ so by induction  
 the pair $(\alpha(e'), \alpha(f'))$ is in $u$'s list.  Because $P$ is angle-monotone, $\alpha(u,v)$ is within $90^\circ$ of $\alpha(e')$ and $\alpha(f')$.  Thus the update step applies.  During the update step we add the angle $\alpha(u,v)$ to the pair $(\alpha(e'), \alpha(f'))$, which gives the pair $(\alpha(e), \alpha(f))$.  Thus we add the pair $(\alpha(e), \alpha(f))$ to $v$'s list.

For the ``if'' direction, suppose that the pair $(\alpha(e), \alpha(f))$ is in $v$'s list.  This pair was added to $v$'s list because of an update from some vertex $u$ closer to $s$ applied to some pair $(\alpha(e'), \alpha(f'))$ in $u$'s list.  By induction, there exists an angle-monotone path $P$ from $s$ to $u$ with extreme edges $e'$ and $f'$, and because the update is only performed when $\alpha(u,v)$ is within $90^\circ$ degrees of $\alpha(e')$ and $\alpha(f')$ therefore the edge $(u,v)$ can be added to $P$ to produce an angle-monotone path with extreme edges $e$ and $f$.
\qed
\end{proof}

To improve the efficiency of the algorithm we observe that  it is redundant to store at a vertex $v$ a pair whose wedge contains the wedge of another pair.  
%
%
Therefore, we only need to store $O(m)$ pairs at each vertex, at most one pair whose first element is $\alpha(e)$ for each edge $e$.  We can simply keep with each vertex $v$ a vector indexed by edges $e$, in which we store the minimal pair $(\alpha(e), \alpha(f))$ (if any) associated with $v$ so far.  
Finally, observe that during the course of the algorithm, each edge $(u,v)$ is handled once in an update step.  
With the refinement just mentioned, handling an edge costs $O(m)$. 
Therefore the algorithm runs in time $O(m^2)$ for a single choice of $s$, and in time $O(n m^2)$ overall. 

The algorithm can be generalized to recognize angle-monotone graphs of width $\gamma$ for fixed $\gamma < 180^\circ$.  It is no longer legitimate to explore vertices in order of distance from $s$, since a generalized angle-monotone path will not necessarily respect this ordering. 
However, we can run the algorithm in phases, where phase $i$ captures all the angle-monotone paths of width $\gamma$ that start at $s$ and have at most $i$ edges.  Since no angle-monotone path can repeat a vertex, there are at most $n-1$ edges in any angle-monotone path.  Thus we need $n-1$ phases.
In each phase, for each directed edge $(u,v)$ we update each pair $(\alpha(e), \alpha(f))$ stored at $u$ as follows.
If $\alpha(u,v)$ is within $\gamma$ of $\alpha(e)$ and within $\gamma$ of $\alpha(f)$ then add to $v$'s list the pair $(\min\{ \alpha(u,v), \alpha(e)\}, \max\{\alpha(u,v), \alpha(f)\})$.
In this way, each of the $n - 1$ phases takes time $O(m^2)$, so the total run-time of the algorithm over all choices of $s$ becomes $O(n^2 m^2)$.

%% file: half-theta-6.tex


In this section we show that every point set in the plane has a plane geometric graph that is angle-monotone with width $120^\circ$.   In particular, we will prove that the \emph{\hts } has this property. 
As noted in the Introduction, there are point sets for which no plane graph is angle-monotone with width $90^\circ$.   It is an open question to narrow this gap and find the minimum
angle $\gamma$ for which every point set has a plane graph 
 that is angle-monotone with width $\gamma$ (and thus spanning ratio $1/ \cos{\gamma \over 2}$).

\begin{figure}[htb]
\centering
\includegraphics[width=1.0\textwidth]{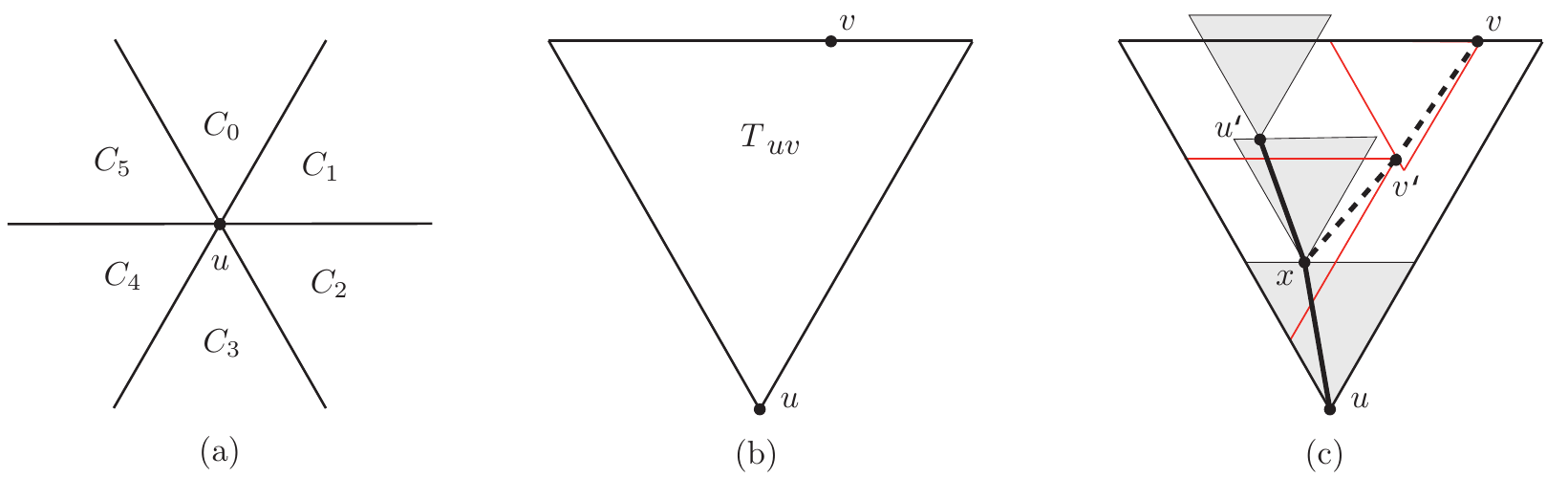}
\caption{(a) 6 cones originating from point $u$, (b) Canonical triangle $T_{uv}$,
\newstuff{(c) path $\sigma_u$ (solid) with its empty canonical triangles shaded, path $\sigma_v$ (dashed) and their common vertex $x$.}}
\label{fig:conecan}
\end{figure}

We first define the \hts.  For each point $u \in P$, partition the plane into $60^\circ$ cones with apex $u$, with each cone defined by two rays at consecutive multiples of $60^\circ$ from the positive $x$-axis. Label the cones $C_0$, $C_1$, $C_2$, $C_3$, $C_4$, and $C_5$ in clockwise order around $u$, starting from the cone containing the positive $y$-axis.  
\oldstuff{Let $C^u_i$ denote the cone $C_i$ with apex $u$.}  
See Figure~\ref{fig:conecan}(a).

For two vertices $u$ and $v$ the \emph{canonical triangle} $T_{uv}$ is the triangle bounded by: the cone of $u$ that contains $v$; and the line through $v$ perpendicular to the bisector of that cone. See Figure~\ref{fig:conecan}(b). Notice that if $v$ is in an even cone of $u$, then $u$ is in an odd cone of $v$. We build the \hts as follows. 
\changed{For each vertex $u$ and each even $i=0,2,4$, add the edge $uv$ provided that $v$ is in the $C_i$ cone of $u$ and $T_{uv}$ is empty.  We call $v$ the \emph{$C_i$-neighbour of $u$}.}
For simplicity, we assume that no two points lie on a line parallel to a cone boundary, guaranteeing that each vertex connects to exactly one vertex in each even cone. Hence the graph has at most $3n$ edges in total. 
The \hts is a type of Delaunay triangulation where the empty region is an equilateral triangle in a fixed orientation as opposed to a disk~\cite{bonichonTD}.   
\changed{It can be computed in $O(n \log n)$ time~\cite{Spanners}.}

\newstuff{To prove angle-monotonicity properties of the \hts, we use an idea like the one used by Angelini~\cite{Angelini}.  His goal was to show that every abstract triangulation has an embedding that is monotone, i.e.~angle-monotone with width $180^\circ$. (The same result was obtained in~\cite{HeHe} with a different proof.)   Angelini did this by showing that the Schnyder drawing of any triangulation is monotone, and in fact, upon careful reading, his proof shows that any Schnyder drawing is angle-monotone with the smaller width $120^\circ$.  Schnyder drawings are a special case of \hts{s}~\cite{bonichonTD} so it is not surprising that Angelini's proof idea extends to the \hts in general.

\begin{theorem} \label{thm:anglemonotone}
The \hts  is angle-monotone with width $120^\circ$.
\end{theorem}
\begin{proof}
We must prove that for any points $u$ and $v$, there is an angle-monotone path from $u$ to $v$ of width $120^\circ$.  Assume without loss of generality that $v$ is in the $C_0$ cone of $u$.  See Figure~\ref{fig:conecan}(b). 

Our path from $u$ to $v$ will be the union of two paths, each of which is angle-monotone of width $60^\circ$.  
We begin by constructing a path $\sigma_u$ from $u$ in which each vertex is joined to its $C_0$ neighbour.  This is a $\beta$-monotone path of width $60^\circ$ for $\beta = 90^\circ$.  If the path contains $v$ we are done, 
\changed{so assume otherwise}.  
Let $u'$ be the last vertex of the path that lies in $T_{uv}$.  Note that $v$ cannot lie in the $C_0$ cone of $u'$.  
Let $S$ be the 
subpath of $\sigma_u$ from $u$ to $u'$, together with the $C_0$ cone of $u'$.  Then $S$ separates $T_{uv}$ into two parts.  Suppose that $v$ lies in the right-hand part (the other case is symmetric).  See Figure~\ref{fig:conecan}(c).

Next, construct a path $\sigma_v$ from $v$ in which each vertex is  joined to its $C_4$ neighbour.  This is a $\beta$-monotone path of width $60^\circ$ for $\beta = 210^\circ$.  

We now claim that $\sigma_u$ and $\sigma_v$ have a common vertex $x$.  Then as our final path from $u$ to $v$ we take the portion of $\sigma_u$ from $u$ to $x$ followed by the portion of $\sigma_v$ backwards from $x$ to $v$.  Since the reverse of $\sigma_v$ is $\beta$-monotone with width $60^\circ$ for $\beta = 30^\circ$, the final path is $\beta$-monotone with width $120^\circ$ for $\beta= 60^\circ$.     

It remains to prove that $x$ exists.  
Let $v'$ be the last vertex of $\sigma_v$ that lies strictly to the right of $S$.  
\changed{Let $u''$ be the last vertex of $\sigma_u$ that lies below $v'$.  We claim that $u''$ is the $C_4$ neighbour of $v'$, and thus that $u''$ provides our vertex $x$. 
Let $T$ be the empty canonical triangle from $u''$ to its $C_0$-neighbour (or the empty $C_0$ cone of $u''$ in case $u''$ has no $C_0$-neighbour).     
First note that $u''$ is in the $C_4$ cone of $v'$---otherwise $v'$ would be in $T$.
Next note that $T_{v'u''}$ is empty---otherwise $vÕ$ would have a $C_4$-neighbour that is in $T$ or is to the right of $S$.}
%
%
\qed 
\end{proof}
}

\oldstuff{\note{Here is our old proof.}
In order to prove angle-monotonicity properties of the \hts, we need the following observation.

\begin{observation} \label{obs:reverse}
If  the path $v_0, v_1, \ldots, v_n$ is $\beta$-monotone with width $\gamma$, then the reverse path 
$v_n, v_{n-1}, \ldots, v_0$ is $(180^\circ - \beta)$-monotone with width $\gamma$.
\end{observation}


\begin{theorem} \label{thm:anglemonotone}
The \hts  is angle-monotone with width $120^\circ$.
\end{theorem}

\begin{proof}
We must prove that for any points $u$ and $v$, there is an angle-monotone path from $u$ to $v$ of width $120^\circ$.  Assume without loss of generality that $v$ is in $C^u_0$.  Let $A_{uv} = T_{uv} \cap C^v_4$ and let $B_{uv}= T_{uv} \cap C^v_2$.   See Figure~\ref{fig:conecan}(b).  

We will prove by induction on the number of points in $T_{uv}$ that there is 
a $uv$ path that is $\beta$-monotone with width $120^\circ$ for $\beta = 60^\circ$ or $120^\circ$.  Furthermore, if $A_{uv}$ is empty then there is a path with $\beta=120^\circ$, and if $B_{uv}$ is empty then there is a path with $\beta=60^\circ$.

{\bf Base Case:} $T_{uv}$ is empty of points. Then $uv$ is an edge of the \hts, which gives the result immediately.

{\bf Inductive Step:}
Assume that $T_{uv}$ is not empty and that the inductive hypothesis holds for all canonical triangles with fewer points than $T_{uv}$.  Let $w$ be the point in $T_{uv}$ such that $T_{uw}$ is empty.
We consider three cases depending on which cone of $v$ contains $w$.

{\bf Case (a):}  $w$ is in $C^v_3$.  Then $v$ is in $C^w_0$.  Note that the canonical triangle $T_{wv}$ has fewer points.  Thus, by induction, there is a $\beta$-monotone path $P$ from $w$ to $v$ of width $120^\circ$, with $\beta = 60^\circ$ or $120^\circ$.  For either value of $\beta$, the edge $uw$ can be added to the beginning of $P$ to 
give a $\beta$-monotone path from $u$ to $v$ of width $120^\circ$, with the same $\beta$.
We just need to check the conditions when $A_{uv}$ or $B_{uv}$ is empty.  So suppose $A_{uv}$ is empty (the case for $B_{uv}$ is similar).  Observe that $A_{w,v}$ is a subset of $A_{uv}$, and therefore is also empty.  Thus, by induction, $P$ can be chosen to be a $120^\circ$-monotone path, and the final $uv$ path is also a $120^\circ$-monotone path.

{\bf Case (b):} $w$ is in $C^v_4$.  Then $w$ is in $A_{uv}$.  We will show that there is a $60^\circ$-monotone $uv$ path with width $120^\circ$.  Observe that this will satisfy all our requirements: our path has the correct angle in case $B_{uv}$ is empty, and $A_{uv}$ cannot be empty since $w$ is in it.

Observe that the canonical triangle $T_{vw}$ has fewer points than $T_{uv}$ since it is a subset.  Also observe that $A_{v,w}$  is empty, since it is a subset of $T_{uw}$. See Figure~\ref{fig:conecan}(c).
By induction, there is a $240^\circ$-monotone path $P$ from $v$ to $w$ with width $120^\circ$.
By Observation \ref{obs:reverse}, $\bar P$, the reverse of $P$, is a $60^\circ$-monotone path from $w$ to $v$ with width $120^\circ$.  The edge $uw$ can be added to the beginning of $\bar P$ to give a  $60^\circ$-monotone path from $u$ to $v$ with width $120^\circ$, as required.

{\bf Case (c):} $w$ is in $C^v_2$.  This is symmetric to the previous case.
\qed
\end{proof}

We note that Theorem~\ref{thm:anglemonotone} implies the results of He and He~\cite{HeHe}, and Angelini~\cite{Angelini} that any abstract triangulation has a planar straight-line embedding that is monotone, i.e.~angle-monotone with width $180^\circ$.  The Schnyder drawing of an abstract triangulation is a \hts~\cite{bonichonTD}, which by Theorem~\ref{thm:anglemonotone} is angle-monotone with width $120^\circ$, and thus monotone.  
A careful reading of~\cite{HeHe} and~\cite{Angelini} shows that they in fact prove that Schnyder drawings are angle-monotone with widths $135^\circ$ (resp.~$120^\circ$), but their results do not imply ours since they do not construct a geometric graph on an arbitrary point set.
}

Theorem \ref{thm:anglemonotone} implies that the spanning ratio of the \hts  is 2, which was already known~\cite{Chew89}.  
The best routing ratio achievable for the \hts is $5/\sqrt{3} \approx 2.887$~\cite{BoseTheta6journal}.  (This was the first \changed{proved separation between spanning ratio and routing ratio}.)
Since angle-monotone paths of width $120^\circ$ have spanning ratio $2$, this implies that 
no local routing algorithm can compute angle-monotone paths with width $120^\circ$ on the \hts.

%


%% file: routing.tex
In this section we give a simple local ``angle'' routing algorithm that finds a path from $s$ to $t$ in any triangulation.  Like previous algorithms, the path walks only along edges of triangles that intersect the line segment $st$. 
The novelty is that the next edge of the path is chosen 
based on angles relative to the vector $st$. 

The details of the algorithm are in 
Section~\ref{sec:paz_routing}.  In Section~\ref{sec:analysis} we prove that the algorithm has routing ratio $1 + \sqrt 2$ on Gabriel graphs, and discuss its behaviour on 
Delaunay triangulations. 
In Section~\ref{sec:limit_routing} we give lower bounds on the routing and competitive ratios of local routing algorithms on Gabriel graphs.

\subsection{Local Angle Routing}
\label{sec:paz_routing}

\begin{figure}[htb]
\centering
\includegraphics[width=0.85\textwidth]{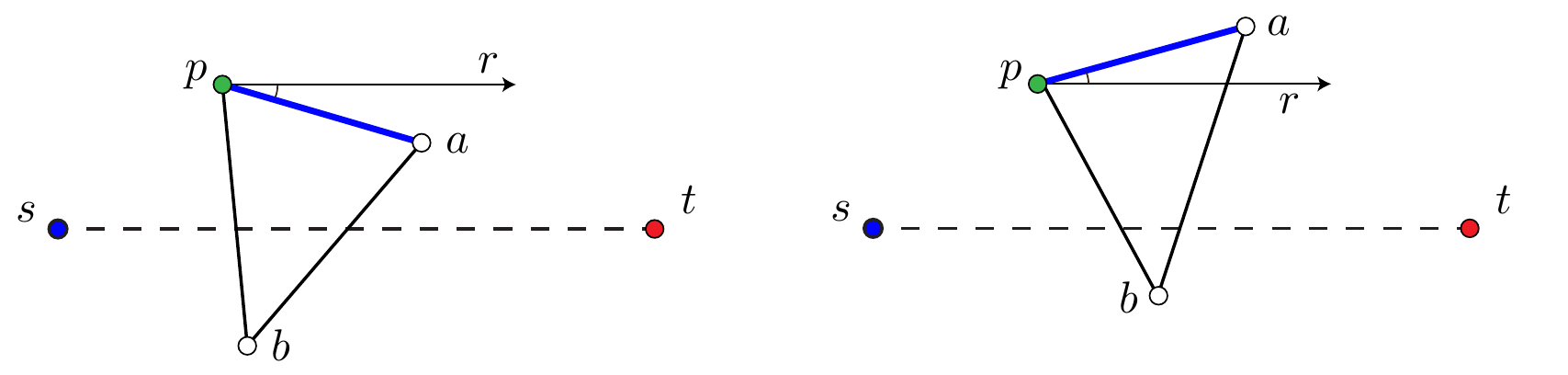}
\caption{Local routing from $s$ to $t$.  At vertex $p$, with $pab$ being the rightmost triangle incident to $p$ that intersects line segment $st$, we route from $p$ to $a$ because the (unsigned) angle $apr$ is less than angle $bpr$.  A step of type $A$ is shown on the left and a step of type $B$ on the right.
}
\label{fig:routing-alg}
\end{figure}

\begin{figure}[htb]
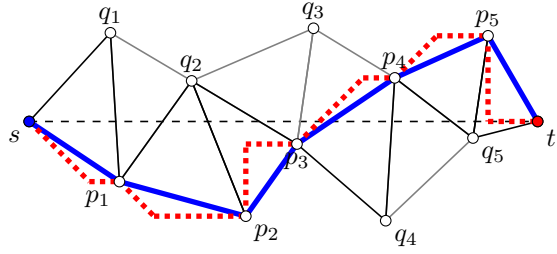

\centering
\figexampleRouting
\caption{Example of route computed by Algorithm~\ref{algo:local-route} (heavy blue path). In dotted red, a longer route obtained by replacing each segment of the route by the most extreme angle.  Both routes are within $(1 + \sqrt 2 )$ of $||st||$.}
\label{fig:route-example}
\end{figure}

Our algorithm is simple to describe:  Suppose we want a route from $s$ to $t$ in a triangulation.  Orient $st$ horizontally, $t$ to the right.  
Suppose we have reached vertex $p$. 
Consider the last (rightmost) triangle that is incident to $p$ and intersects the line segment $st$.
The triangle has two edges incident to $p$.  Of these two edges, take the one that has the minimum angle to the horizontal ray from $p$ to the right.    See Figure~\ref{fig:routing-alg}.
Pseudo-code can be found below in Algorithm~\ref{algo:local-route}.  
\changed{Note that in the pseudo-code, the angle test is equivalently replaced by two tests, identifying steps of type A and B for easier case analysis.}
\changed{For an example of a path computed by the algorithm, see Figure~\ref{fig:route-example}.}
Observe that the algorithm always succeeds in finding a route from $s$ to $t$ because it always advances rightward in the sequence of triangles that intersect line segment $st$.  

%

%



\begin{algorithm}[htb]
\DontPrintSemicolon
$p \leftarrow s$\\
\While{$p \ne t$}{ 
Let $T=pab$ be the rightmost triangle containing $p$ that intersects segment $st$, with $p$ and $a$ on the same side of line $st$.\\
\If(\tcc*[f]{step of type $A$}){
$a$ is closer to line $st$ than $p$
}{
  $p \leftarrow a$ 
}\Else(\tcc*[f]{step of type $B$}){ 
  \If{$| {\it slope} (pa)| \leq | {\it slope} (pb)|$}{
$p \leftarrow a$ 
}
\Else{
$p \leftarrow b$ 
}
}
}
\caption{Local angle routing}
\label{algo:local-route}
\end{algorithm}

\subsection{Analysis of the Algorithm}
\label{sec:analysis}

In this section we will prove that the above algorithm has routing ratio exactly $1 + \sqrt 2$ on Gabriel triangulations,
which have maximum angle  at most $90^\circ$.
In the last part of the section we generalize the analysis to triangulations with a larger maximum angle, and we 
show that the routing ratio is at least 5.07 on Delaunay triangulations.

The intuition for bounding the routing ratio on Gabriel triangulations is to replace each segment of the route by the most extreme segment possible.  See Figure~\ref{fig:route-example}. Any step of type B is replaced by a $45^\circ$ segment plus a horizontal segment.  Any step of type A is replaced by a vertical segment plus a horizontal segment. 
Vertical segments are the bad ones, but each vertical must be preceded by  $45^\circ$ segments, which means that instead of travelling 1 unit horizontally (the optimum route) we have travelled $\sqrt 2$ along a $45^\circ$ segment plus 1 vertically, giving us the  $1 + \sqrt 2$ ratio.  We now give a more formal proof.

For each edge $e = (p_i, p_{i+1})$ of the path, let $d_x(e) = ||x(p_i)-x(q_{p+1})||$ and $d_y(e) = ||y(p_i)-y(p_{i+1})||$. 
Let $A$ (resp.~$B$) be the set of edges of the path where the algorithm makes a step of type $A$ (resp.~type $B$). 
(Context will distinguish edge sets from steps.) 
Let $x_B = \sum_{e\in B} d_x(e)$ and $x_A = \sum_{e \in A} d_x(e)$.

\begin{lemma}
\label{lem:increasing}
\changed{On any Gabriel triangulation} the path computed by Algorithm~\ref{algo:local-route} is $x$-increasing. 
\end{lemma}

\begin{proof} 
Let us show that each step is $x$-increasing.  
Consider a step from $p$, with $a$ and $b$ as defined in Algorithm~\ref{algo:local-route}.
Assume without loss of generality that $p$ and $a$ are above line $st$ and $b$ is below. Since $T$ is the last triangle incident to $p$ that intersects $st$, 
the clockwise ordering of $T$ is $pab$.  Refer to Figure~\ref{fig:routing-alg}.

If the algorithm takes a step of type $B$ then $a$ is above $p$ (in $y$ coordinate) and $b$ is below $p$.  Since 
$\angle bpa \leq 90^\circ$, thus $x(a)$ and $x(b)$ are greater than $x(p)$.  
If the algorithm takes a step of type $A$ then since $b$ is below $st$ and $a$ is above $st$ and $\angle bap \leq 90^\circ$, thus $x(a)$ is greater than $x(p)$.
\qed
\end{proof}

\remove{
\begin{itemize} 
\item case 1: $(qq')$ is a step $B$ edge.  Since $q''$ is below $(st)$ and the angle $(q”qq') \leq 90^\circ, x(q')>x(q)$.
\item case 2: $(qq')$ is a step $A$ edge.  Since $q''$ is below (st) and the angle $(qq'q'') \leq 90^\circ, x(q')> x(q)$.
\item case 3: $(qq'')$ is a step $B$ edge. Since $q'$ is above q and the angle $(q'qq”) \leq 90^\circ, x(q”)>x(q)$.
\end{itemize}
}

\begin{theorem}
\label{thm:routing_paz} On any Gabriel triangulation, 
Algorithm~\ref{algo:local-route}
has a routing ratio of $1+\sqrt{2}$ and this bound is tight.
\end{theorem}

\begin{proof}
We first bound $\sum_{e \in B} ||e||$.
Observe that each edge in $B$ forms an angle with the horizontal line through $p$ that is at most $45^\circ$.
Thus $\sum_{e \in B} d_y(e) \leq x_B$ and $\sum_{e \in B} ||e|| \leq \sqrt{2} x_B$.

We next bound $\sum_{e \in A} ||e||$.
Observe that edges in $A$ move us closer to the line $st$, and must be balanced by previous steps (of type $B$) that moved us farther from the line $st$.
This implies that $\sum_{e \in A} d_y(e) \le \sum_{e \in B} d_y(e) \leq x_B$ (where the last step comes from the first observation).
Since $||e|| \le d_x(e) + d_y(e)$, thus $\sum_{e \in A} ||e|| \leq x_A + \sum_{e \in A} d_y(e) \leq x_A + x_B$.

Putting these together, the length of the path is bounded by 
$\sum_{e \in A} ||e|| + \sum_{e \in B} ||e|| \leq x_A + x_B + \sqrt{2} x_B \leq (1 + \sqrt 2)(x_A + x_B)$.
Finally, by Lemma~\ref{lem:increasing}, $x_A + x_B = ||st||$, so this proves that the routing ratio is at most $(1 + \sqrt 2)$.

\remove{
Observation 2: only type A steps can increase the distance of the traveler with the line $(st)$.  
Hence:
$\sum_{e \in A} dy(e) <= \sum_{e \in B} dy(e) \leq x_B.$
Since $||e|| <= dy(e) + dx(e)$, $\sum_{e \in B} ||e||\leq x_B + x_A$.
Putting things together gives us (and using Lemma~\ref{lem:increasing}) we get:
$\sum_{e \in A} ||e|| + \sum_{e \in B} ||e|| \leq \sqrt{2} x_A  + x_B + x_A \leq (1+\sqrt{2})||st||$.
}

An example to show that this analysis is tight is given in 
Appendix~\ref{sec:appendix-routing}.
\qed
\end{proof}


We conclude this section with two results on the behaviour of the routing algorithm on other triangulations.
Proofs 
are deferred to Appendix~\ref{sec:appendix-routing}. 

\begin{theorem}
\label{thm:routing_general_angle}
In a triangulation with maximum angle $\alpha < 120^\circ$ Algorithm~\ref{algo:local-route} has a routing ratio of $(\sin \alpha + \sin {\alpha \over 2}) / \sin {{3 \alpha} \over 2}$ and this bound is tight.
\end{theorem}

\begin{theorem}
\label{thm:Del-lower-bound}
The routing ratio of Algorithm~\ref{algo:local-route} on Delaunay triangulation is greater than $5.07$.
\end{theorem}

We believe \changed{that the routing ratio of Algorithm~\ref{algo:local-route} on Delaunay triangulations is close to $5.07$, but leave that as an open question.}
We remark that Algorithm~\ref{algo:local-route} is different from  the generalization of Chew's Routing Algorithm for Delaunay triangulations~\cite{Chew86}  (cf.~the algorithm described in~\cite{BBCPR15}).  


\remove{ 
\subsection{Performance of the Algorithm on Delaunay Triangulations}
\label{sec:generalisation}

\bluenote{This section still needs proofreading.}

Remark: the previous routing algorithm can also be used in classical Delaunay triangulations. 
First let us remark that the proposed algorithm is different from the the generalization of Chew's Routing~\cite{Chew86} Algorithm for Delaunay triangulations (cf. algorithm described in~\cite{BBCPR15}).

{\bf Open question 1}: Is it possible to bound the routing ratio of this algorithm on classical Delaunay Triangulations ? If so, what is the best upper bound ?

We can first proposed a lower bound on its routing ratio:

\begin{theorem}
The routing ratio of the proposed algorithm on Delaunay triangulation is greater than $5.07$.
\end{theorem}
\begin{proof}(\emph{very sketchy})

\begin{figure}
\centering\includegraphics[width=0.5\textwidth]{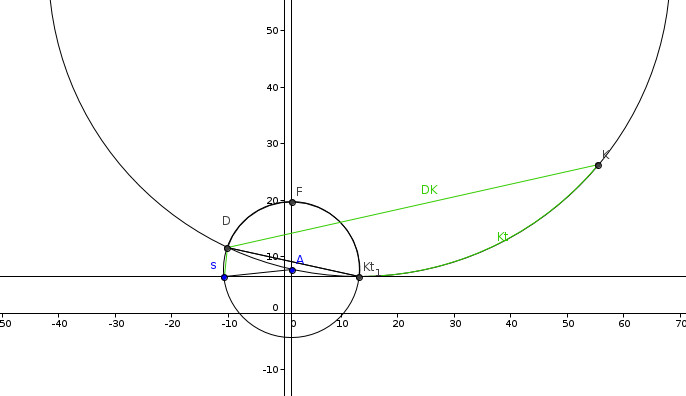}
\caption{Example that gives a 5.05 lower bound on the routing ratio of the proposed routing algorithm on triangulations (\texttt{./otherMaterial/LB\_low\_stretch.ggb}).}
\label{fig:4}
\end{figure}

Figure~\ref{fig:4} gives a readable example with routing ratio of 5.50. It's possible to tune it in order to obtain a routing ratio of 5.07.
Let us explain this example.
The first triangle $(sDD')$ is such that line $(sD')$ is almost tangent to the first circle at s and that the slope of line $(sD)$ is slightly smaller than the slope $(sD')$, so we route to $D$.
Once at $D$ the triangle $(DKK')$ is such that $K'$ is very close to $t$ and $(tK)$ is a slope smaller than $(tK')$. So we route to $K$. Then we complete the figure with
a fan of triangles that are all connected to $K'$. So we route from $K$ to $t$ using the arc.
In the picture the computed route is in green.
When $s$ and $D$ are getting closer, the routing ratio seems to converge toward something close to 5.07.
\end{proof}

{\bf Open question 2}: can we express the routing ratio of this algorithm with respect to the maximum angle in any triangle (e.g. $90^\circ$ for Gabriel triangulations and $180^\circ$ for Delaunay triangulation). For maximal angle less than $120^\circ$ the proof of Theorem~\ref{thm:routing_paz} can be adapted leading to a routing ratio that is no longer tight (even unbounded when the maximal angle equals $120^\circ$.
}

\subsection{Limits of Local Routing Algorithms on Gabriel Triangulations}
\label{sec:limit_routing}

In this section we prove some limits on local routing on Gabriel triangulations.  Proofs are deferred to 
Appendix~\ref{sec:appendix-routing}. 

\changed{
A routing algorithm on a geometric graph $G$ has a \emph{competitive ratio}
of $c$ if the length of the path produced by the algorithm from any vertex $s$
to any vertex $t$ is at most $c$ times the length of the shortest path from
$s$ to $t$ in $G$, \changed{and $c$ is the minimum such value}. 
(Recall that the routing ratio compares the length of the path produced by the algorithm to the Euclidean distance between the endpoints.  Thus the competitive ratio is less than or equal to the routing ratio.)

A routing algorithm is 
\emph{$k$-local} (for some integer constant $k>0$)
if it makes forwarding decisions based on: (1) the $k$-neighborhood in $G$  of the current position of the message; and (2) limited information stored in the message header.
}

\begin{theorem}
\label{thm:lb_routing_gabriel} 
Any $k$-local routing algorithm on Gabriel triangulations has routing ratio at least 1.4966 and competitive ratio at least 1.2687.
%
\end{theorem}

Although Gabriel triangulations are angle-monotone~\cite{D-Frati-G},
Theorem~\ref{thm:lb_routing_gabriel} shows that no local routing algorithm can compute angle-monotone paths since that would give routing ratio $\sqrt{2}$. The following theorem tells us that even less constrained paths cannot be computed locally:

\begin{theorem}
\label{thm:no_local_self_approaching}
There is no $k$-local routing algorithm on Gabriel triangulations that always finds self-approaching paths.
\end{theorem}

%% file: conclusions.tex

\remove{An implementation of Algorithm~\ref{algo:local-route} (working on any Delaunay triangulations) is available at the following address: \url{http://www.labri.fr/~bonichon/routing/routing.jar}.}

We conclude this paper with some open questions. 

\begin{enumerate}
\item What is the minimum
angle $\gamma$ for which every point set has a plane geometric graph 
 that is angle-monotone with width $\gamma$ (and thus has spanning ratio $1/ \cos{\gamma \over 2}$)?  We proved $\gamma \le 120^\circ$, and it is known that $\gamma > 90^\circ$.

\item Is there a local routing algorithm with bounded routing ratio for any angle-monotone graph?  \changed{Any} increasing-chord graph?   

\item We bounded the routing ratio of our local routing algorithm on triangulations  based on the maximum angle in the triangulation, but how does this relate to the property of being generalized angle-monotone?  
If a triangulation has bounded maximum angle, is it generalized angle-monotone?  
The only thing known is  that maximum angle $90^\circ$ implies angle-monotone with width $90^\circ$~\cite{D-Frati-G}.

\item Is the standard Delaunay triangulation generalized angle-monotone?  
In particular, proving that the Delaunay triangulation is angle-monotone with width strictly less than $120^\circ$ would provide a different proof that the Delaunay triangulation has spanning ratio less than 2~\cite{xia2013stretch}.  It is known that the Delaunay triangulation is not angle-monotone with width $90^\circ$ (see Section~\ref{sec:introduction}).

\item How does our local routing algorithm behave on standard Delaunay triangulations?  We proved a lower bound of 5.07 on the routing ratio.  We believe the  routing ratio is close to this value, but have no upper bound.

\end{enumerate}

%% file: appendix.tex
\section{Omitted Proofs for Section~\ref{sec:introduction}}
\label{sec:appendix-intro}

\begin{proof}[Proof of Observation~\ref{obs:ratio}]
In the worst case we travel the two equal sides of an isoceles triangle with base length 1 and two angles of $\gamma /2$.  If $\ell$ is the side length, the ratio is $2\ell$, and we have $\cos {\gamma \over 2} = {1 \over 2} / \ell$.  Thus the ratio is $1 / \cos {\gamma \over 2}$.
\qed
\end{proof}

\begin{figure}
\center{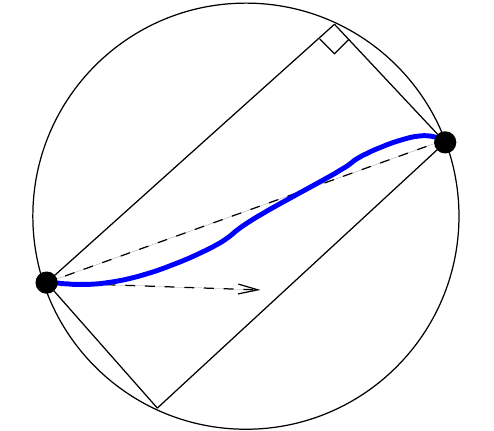}
\caption{Illustration of Lemma~\ref{lemma:disc}. A $\beta$-monotone path (in blue) inside the rectangle with both sides at angles $\beta \pm 45^\circ$. This rectangle lies inside the disc of diameter $uv$.}
\label{fig:lemma1}
\end{figure}

\section{Omitted Proofs for Section~\ref{sec:routing}}
\label{sec:appendix-routing}

\begin{figure}
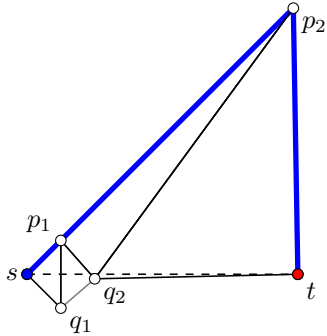

\centering
\figLBAlgo
\caption{Example that gives a lower bound on the routing ratio of our routing algorithm on Gabriel triangulations.  The route found by the algorithm is drawn as a heavy blue path.}
\label{fig:LBAlgo}
\end{figure}

\begin{proof}[Proof of Theorem~\ref{thm:routing_paz}]
To complete the proof, we show an example for which our algorithm gives a routing ratio of $1+\sqrt{2}$.
Consider the configuration shown in Figure~\ref{fig:LBAlgo}. It is a Gabriel triangulation and the route computed by the algorithm is as shown. Observe that the size of the leftmost circle can be made arbitrarily small compared to $||st||$. Hence, when $s=(0,0)$ and $t=(1,0)$, the route can be arbitrarily close to the polyline $s \rightarrow (1,1) \rightarrow t$. Thus we can build a point set such that the length of the computed route is as close to $1+\sqrt{2}$ as we want.
\qed
\end{proof}

\begin{figure}[htb]
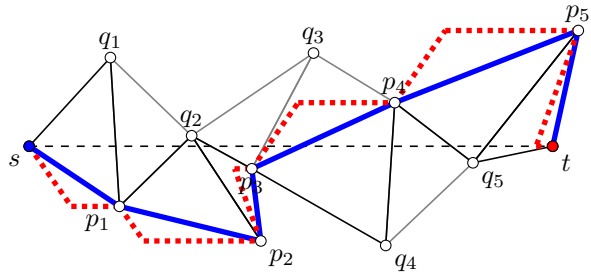

\centering
\figexampleRoutingGeneralized
\caption{Intuition for general routing.}
\label{fig:general-route-intuition}
\end{figure}

\begin{figure}[htb]
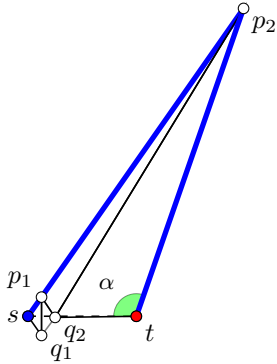

\centering
\figLBAlgoGeneralized
\caption{The worst case situation for general routing.}
\label{fig:general-route-worst}
\end{figure}

\begin{proof}[Proof sketch for Theorem~\ref{thm:routing_general_angle}]
Following the intuitive justification for the routing ratio of Algorithm~\ref{algo:local-route}
on Gabriel triangulations, lengthen the route by replacing each segment of the route by the most  extreme segment possible. 
Any step of type B is replaced by a segment at angle $\alpha \over 2$ plus a horizontal segment.  Any step of type A is replaced by a segment at angle $\alpha$ 
plus a horizontal segment.  In all cases angles are measured from the forward horizontal.
See Figure~\ref{fig:general-route-intuition}.
Segments of type A are the bad ones, but each such segment must be preceded by angle $\alpha \over 2$ segments, which means that instead of travelling 1 unit horizontally (the optimum route) we have travelled on a segment of angle $\alpha \over 2$ and then on a segment of angle $- \alpha$ (both angles measured w.r.t~the forward horizontal).  Let these segments have lengths $\ell_1$ and $\ell_2$ respectively.  In the triangle formed by these three segments, the $\ell_1$ segment is opposite angle $\alpha$, the $\ell_2$ segment is opposite angle ${\alpha \over 2}$ and the unit horizontal is opposite angle $180^\circ - {{3 \alpha} \over 2}$.  
We need $180^\circ - {{3 \alpha} \over 2} >0$, i.e.~$\alpha < 120^\circ$.
By the sine law, $\ell_1 = \sin \alpha / \sin {{3 \alpha} \over 2}$ and $\ell_2 = \sin {\alpha \over 2} / \sin {{3 \alpha} \over 2}$.  Thus  the distance travelled is $\ell_1 + \ell_2 = (\sin \alpha + \sin {\alpha \over 2}) / \sin {{3 \alpha} \over 2}$.

To show that the bound is tight, we generalize the example of Figure~\ref{fig:LBAlgo}. The resulting example is shown in Figure~\ref{fig:general-route-worst}.
\qed
\end{proof}

\begin{figure}[htb]
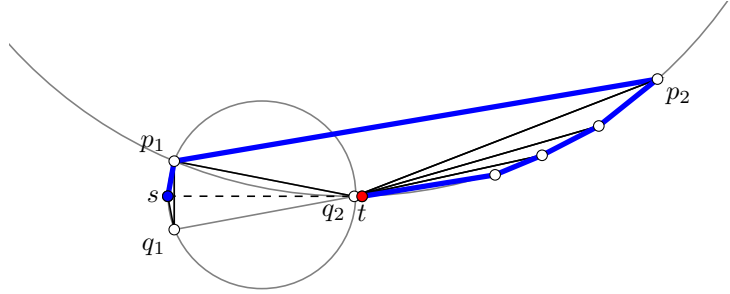

\centering
\figLBDel
\caption{Example that gives a 5.0 lower bound on the routing ratio of Algorithm~\ref{algo:local-route} on Delaunay triangulations.}
\label{fig:4}
\end{figure}

\begin{proof}[Proof of Theorem~\ref{thm:Del-lower-bound}]
Let us explain the example of Figure~\ref{fig:4}. This Delaunay triangulation is defined in the following way:
The first triangle $sp_1q_1$ is such that the slope of line $sp_1$ is slightly smaller than the slope of $sq_1$, so we route to $p_1$. Let $q_2$ be a point on the empty circle $C_0$ containing $s$, $p_1$ and $q_1$ that is slightly below the $x$-axis. Let $C_1$ be the circle that goes through $p_1$ and $q_2$ such that the tangent of $C_1$ at $q_2$ is horizontal. Let $p_2$ be a point on $C_1$ such that the slope of $p_1p_2$ is slightly smaller than the slope of $p_1q_2$. We place point $t$ at the rightmost intersection of $C_1$ and the $x$-axis, and we place vertices densely on the arc of $C_1$ between $p_2$ and $t$.
The route in the example of Figure~\ref{fig:4} has length about $5||st||$. Moving $p_1$ closer and closer to $s$ leads to $5.07$ as a lower bound on the routing ratio of Algorithm~\ref{algo:local-route} on Delaunay triangulations.
\qed
\end{proof}

\remove{
A routing algorithm on a geometric graph $G$ has a \emph{competitive ratio}
of $c$ if the length of the path produced by the algorithm from any vertex $s$
to any vertex $t$ is at most $c$ times the length of the shortest path from
$s$ to $t$ in $G$, \changed{and $c$ is the minimum such value}. 
(Recall that the routing ratio compares the length of the path produced by the algorithm to the Euclidean distance between the endpoints.  Thus the competitive ratio is less than or equal to the routing ratio.)
A routing algorithm is 
\emph{$k$-local} 
if it makes forwarding decisions based on 1) the $k$-neighborhood in $G$ (for some
integer constant $k>0$) of the current position of the message and 2) limited
information stored in the message header.
}

\begin{proof} [Proof of Theorem~\ref{thm:lb_routing_gabriel}]

\begin{figure}[htb]
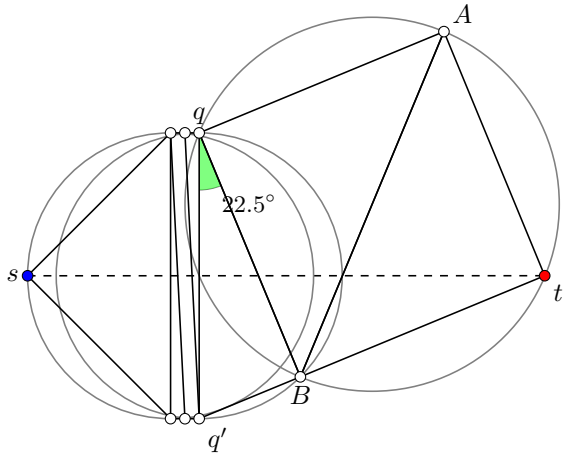

\centering
\figLB
\caption{Example for lower bounds on the routing ratio and competitive ratio of any $k$-local routing algorithm on Gabriel triangulations.}
\label{fig:LB}
\end{figure}

Let us consider the triangulation of Figure~\ref{fig:LB}. This triangulation is defined as follows: all the triangles intersecting the segment $st$ are right triangles. The first one is isosceles and symmetric with respect to the $x$-axis. Then we have a \emph{fan} of $2k-2$ triangles each having a horizontal side and 
pointing alternately upward and downward.  Let $q$ and $q'$ be respectively the upper rightmost and lower rightmost points of this set of triangles. The next triangle $q q' B$ is such that  the angle $\angle q' q B= 22.5^\circ$. The point $t$ is on the intersection of the line $q'B$ and the $x$-axis. We complete the triangulation with two triangles, $qBA$ and $ABt$ having common hypotenuse $AB$. Finally we make the fan of of $2k-2$ triangles arbitrarily thin and we assume that $||qq'|| = 2$.

  Now let us consider any deterministic $k$-local routing algorithm. We
consider two triangulations: The first is the one described above
(and shown in Fig.~\ref{fig:LB}) and the second one is obtained from
the first by reflecting over the $x$-axis the part of the triangulation that lies to the right of
$q q'$. No deterministic $k$-local routing algorithm computing
a path from $s$ to $t$ can
distinguish between the two point sets until a vertex less than $k$ hops away
from $q$ or $q'$ is reached. Let $q''$ be the vertex $k$ hops away from $q$ or
$q'$ that is reached by the algorithm on either triangulation.

 Since the  fan is arbitrarily thin, $q''$ can be assumed to be arbitrarily
close to $q$ or to $q'$. 

Each case, $q$ or $q'$, leads to a non-optimal path
for one of the point sets; we only consider the first case as the second will
follow by symmetry. If $q''$ is arbitrarily close to $q$ then, for the
point set shown in Fig.~\ref{fig:LB}, the shortest paths from $q''$ to $t$
go through $A$ or $B$ and are of length $||qB||+||Bt|| = 2 ||qB|| = 4 \cos(22.5^\circ)$. Moreover $||sq|| = \sqrt{2}$ and $||st|| = 1+ 1/\tan(22.5^\circ)$.
Hence the length of the complete path computed by the algorithm is at least $\frac{\sqrt 2+ 4 \cos(22.5^\circ)}{1+ 1/\tan(22.5^\circ)} ||st|| \approx 1.496605761$, which proves the routing ratio lower bound. The shortest route from $s$ to $t$ goes through $q'$ and is of length $||sq'|| + ||q'B||+ ||q't|| = \sqrt 2 +(1-\cos(45^\circ))/\sin(22.5^\circ) + 2\cos(22.5^\circ)$. Thus a  lower bound on the competitive ratio is  $\frac{\sqrt 2+ 4 \cos(22.5^\circ)}{\sqrt 2 + (1-\cos(45^\circ))/\sin(22.5^\circ) + 2\cos(22.5^\circ)} \approx 1.268761101$.
\qed
\end{proof}

 \begin{figure}[htb]
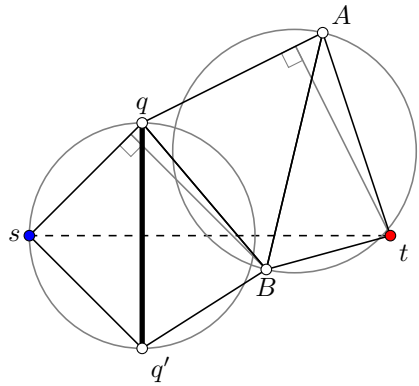

\centering
\figNoSA
\caption{Example of Gabriel triangulation used to show that no $k$-local routing algorithm can compute self-approaching paths.}
\label{fig:NoSA}

\end{figure}

\begin{proof}[Proof sketch for Theorem~\ref{thm:no_local_self_approaching}] 
We apply reasoning as in the previous proof, but this time on the triangulation of Figure~\ref{fig:NoSA}, where the fat segment $qq'$ represents a fan of $2k-2$ thin triangles. As before we assume that the algorithm is routing through $q$ (if not we consider the symmetric triangulation with respect to the $x$-axis).
Moving from $s$ to $q$ the distance toward $B$ is not decreasing. Hence a self approaching path that goes through the edge $sq$ cannot go through the vertex $B$. Hence once at vertex $q$ the only possibility is to use the edge $qA$. But moving along the edge $qA$ the distance toward $t$ is not decreasing. Hence, there is no self approaching path from $s$ to $t$ that goes through $q$.

So for any deterministic $k$-local routing algorithm, there exists a triangulation on which the algorithm will not find a self-approaching path.
\qed
\end{proof}

%% file: figs/lemma1.pdf_t
\begin{picture}(0,0)%
\includegraphics{lemma1.pdf}%
\end{picture}%
\setlength{\unitlength}{3108sp}%
\begingroup\makeatletter\ifx\SetFigFont\undefined%
\gdef\SetFigFont#1#2#3#4#5{%
  \reset@font\fontsize{#1}{#2pt}%
  \fontfamily{#3}\fontseries{#4}\fontshape{#5}%
  \selectfont}%
\fi\endgroup%
\begin{picture}(3000,2614)(2551,-3888)
\put(2566,-3346){\makebox(0,0)[lb]{\smash{{\SetFigFont{9}{10.8}{\rmdefault}{\mddefault}{\updefault}$u$}}}}
\put(5536,-2221){\makebox(0,0)[lb]{\smash{{\SetFigFont{9}{10.8}{\rmdefault}{\mddefault}{\updefault}$v$}}}}
\put(3466,-3256){\makebox(0,0)[lb]{\smash{{\SetFigFont{9}{10.8}{\rmdefault}{\mddefault}{\updefault}$\beta$}}}}
\end{picture}%

%% file: angle-monotone.bbl
\begin{thebibliography}{10}
\providecommand{\url}[1]{\texttt{#1}}
\providecommand{\urlprefix}{URL }

\bibitem{Aichholzer:genSAcurves:2001}
Aichholzer, O., Aurenhammer, F., Icking, C., Klein, R., Langetepe, E., Rote,
  G.: Generalized self-approaching curves. Discrete Applied Mathematics
  109(1--2),  3--24 (2001)

\bibitem{Alamdari2013}
Alamdari, S., Chan, T.M., Grant, E., Lubiw, A., Pathak, V.: Self-approaching
  graphs. In: Didimo, W., Patrignani, M. (eds.) Proc. Graph Drawing (GD), LNCS,
  vol. 7704, pp. 260--271. Springer (2013)

\bibitem{Angelini}
Angelini, P.: Monotone drawings of graphs with few directions. In: 6th Int.
  Conf. Information, Intelligence, Systems and Applications (IISA). pp. 1--6.
  IEEE (2015)

\bibitem{Angelini:MonoDraw:2012}
Angelini, P., Colasante, E., Battista, G.D., Frati, F., Patrignani, M.:
  Monotone drawings of graphs. J. Graph Algorithms Appl.  16(1),  5--35 (2012)

\bibitem{Angelini:2009}
Angelini, P., Frati, F., Grilli, L.: An algorithm to construct greedy drawings
  of triangulations. J. Graph Algorithms Appl.  14(1),  19--51 (2010)

\bibitem{bern1990provably}
Bern, M., Eppstein, D., Gilbert, J.: Provably good mesh generation. In: Proc.
  31st Symp. on Foundations of Computer Science (FOCS). pp. 231--241. IEEE
  (1990)

\bibitem{BBCPR15}
Bonichon, N., Bose, P., De~Carufel, J.L., Perkovi{\'{c}}, L., Van~Renssen, A.:
  Upper and lower bounds for online routing on {D}elaunay triangulations. In:
  Bansal, N., Finocchi, I. (eds.) Proc. 23rd European Symp. on Algorithms
  (ESA). LNCS, vol. 9294, pp. 203--214. Springer (2015)

\bibitem{bonichonTD}
Bonichon, N., Gavoille, C., Hanusse, N., Ilcinkas, D.: Connections between
  theta-graphs, {D}elaunay triangulations, and orthogonal surfaces. In:
  Thilikos, D.M. (ed.) Proc. 36th Int. Workshop Graph Theoretic Concepts in
  Computer Science (WG). LNCS, vol. 6410, pp. 266--278 (2010)

\bibitem{BoseTheta6journal}
Bose, P., Fagerberg, R., van Renssen, A., Verdonschot, S.: Optimal local
  routing on {D}elaunay triangulations defined by empty equilateral triangles.
  {SIAM} J. Comput.  44(6),  1626--1649 (2015)

\bibitem{Chew86}
Chew, L.P.: There is a planar graph almost as good as the complete graph. In:
  Proc. 2nd Annual Symp. Computational Geometry (SoCG). pp. 169--177 (1986)

\bibitem{Chew89}
Chew, L.P.: There are planar graphs almost as good as the complete graph. J.
  Computer and System Sciences  39(2),  205--219 (1989)

\bibitem{D-Frati-G}
Dehkordi, H.R., Frati, F., Gudmundsson, J.: Increasing-chord graphs on point
  sets. J. Graph Algorithms Appl.  19(2),  761--778 (2015)

\bibitem{dumitrescu2015lower}
Dumitrescu, A., Ghosh, A.: Lower bounds on the dilation of plane spanners
  (2015), \url{http://arxiv.org/pdf/1509.07181v3.pdf}

\bibitem{HeHe}
He, X., He, D.: Monotone drawings of 3-connected plane graphs. In: Bansal, N.,
  Finocchi, I. (eds.) Proc. 23rd European Symp. on Algorithms (ESA). LNCS, vol.
  9294, pp. 729--741. Springer (2015)

\bibitem{Icking:self-approachingcurves:1995}
Icking, C., Klein, R., Langetepe, E.: Self-approaching curves. Math. Proc.
  Cambridge Philosophical Society  125,  441--453 (1995)

\bibitem{Leighton:2010}
Leighton, T., Moitra, A.: Some results on greedy embeddings in metric spaces.
  Discrete Comput. Geom.  44,  686--705 (2010)

\bibitem{mulzer}
Mulzer, W.: Minimum Dilation Triangulations for the Regular n-Gon. Master's
  thesis, Freie Universit{\"a}t Berlin (2004)

\bibitem{Spanners}
Narasimhan, G., Smid, M.: Geometric Spanner Networks. Cambridge University
  Press (2007)

\bibitem{Papadimitriou:2005}
Papadimitriou, C.H., Ratajczak, D.: On a conjecture related to geometric
  routing. Theor. Comput. Sci.  344,  3--14 (2005)

\bibitem{Rote:ICcurves:1994}
Rote, G.: Curves with increasing chords. Math. Proc. Cambridge Philosophical
  Society  115,  1--12 (1994)

\bibitem{xia2013stretch}
Xia, G.: The stretch factor of the {D}elaunay triangulation is less than 1.998.
  SIAM J. Comput.  42(4),  1620--1659 (2013)

\end{thebibliography}
